\documentclass[twoside,leqno]{article}
\usepackage{amsmath}
\usepackage{amsthm}
\usepackage{amssymb}
\usepackage{amscd}
\usepackage{graphicx}
\usepackage{epsfig}

\usepackage{latexsym}
\usepackage{amsfonts}
\input xy
\xyoption{all} \tolerance=500

\advance\topmargin by -\headheight\advance\topmargin by -\headsep

\numberwithin{equation}{section} \allowdisplaybreaks

\newtheorem{theorem}{Theorem}[section]

\theoremstyle{definition}
\newtheorem{definition}{Definition}[section]
\newtheorem{example}{Example} [section]

\newtheorem{remark}{Remark}[section]

\begin{document}
\font\black=cmbx10 \font\sblack=cmbx7 \font\ssblack=cmbx5 \font\blackital=cmmib10
\skewchar\blackital='177 \font\sblackital=cmmib7 \skewchar\sblackital='177
\font\ssblackital=cmmib5 \skewchar\ssblackital='177
\font\sanss=cmss10 \font\ssanss=cmss8 
\font\sssanss=cmss8 scaled 600 \font\blackboard=msbm10 \font\sblackboard=msbm7
\font\ssblackboard=msbm5 \font\caligr=eusm10 \font\scaligr=eusm7 \font\sscaligr=eusm5
\font\blackcal=eusb10 \font\fraktur=eufm10 \font\sfraktur=eufm7 \font\ssfraktur=eufm5
\font\blackfrak=eufb10

\font\bsymb=cmsy10 scaled\magstep2
\def\all#1{\setbox0=\hbox{\lower1.5pt\hbox{\bsymb
       \char"38}}\setbox1=\hbox{$_{#1}$} \box0\lower2pt\box1\;}
\def\exi#1{\setbox0=\hbox{\lower1.5pt\hbox{\bsymb \char"39}}
       \setbox1=\hbox{$_{#1}$} \box0\lower2pt\box1\;}

\def\mi#1{{\fam1\relax#1}}
\def\tx#1{{\fam0\relax#1}}

\newfam\bifam
\textfont\bifam=\blackital \scriptfont\bifam=\sblackital
\scriptscriptfont\bifam=\ssblackital
\def\bi#1{{\fam\bifam\relax#1}}

\newfam\blfam
\textfont\blfam=\black \scriptfont\blfam=\sblack \scriptscriptfont\blfam=\ssblack
\def\rbl#1{{\fam\blfam\relax#1}}

\newfam\bbfam
\textfont\bbfam=\blackboard \scriptfont\bbfam=\sblackboard
\scriptscriptfont\bbfam=\ssblackboard
\def\bb#1{{\fam\bbfam\relax#1}}

\newfam\ssfam
\textfont\ssfam=\sanss \scriptfont\ssfam=\ssanss \scriptscriptfont\ssfam=\sssanss
\def\sss#1{{\fam\ssfam\relax#1}}

\newfam\clfam
\textfont\clfam=\caligr \scriptfont\clfam=\scaligr \scriptscriptfont\clfam=\sscaligr
\def\cl#1{{\fam\clfam\relax#1}}

\newfam\frfam
\textfont\frfam=\fraktur \scriptfont\frfam=\sfraktur \scriptscriptfont\frfam=\ssfraktur
\def\fr#1{{\fam\frfam\relax#1}}

\def\cb#1{\hbox{$\fam\gpfam\relax#1\textfont\gpfam=\blackcal$}}

\def\hpb#1{\setbox0=\hbox{${#1}$}
    \copy0 \kern-\wd0 \kern.2pt \box0}
\def\vpb#1{\setbox0=\hbox{${#1}$}
    \copy0 \kern-\wd0 \raise.08pt \box0}

\def\pmb#1{\setbox0\hbox{${#1}$} \copy0 \kern-\wd0 \kern.2pt \box0}
\def\pmbb#1{\setbox0\hbox{${#1}$} \copy0 \kern-\wd0
      \kern.2pt \copy0 \kern-\wd0 \kern.2pt \box0}
\def\pmbbb#1{\setbox0\hbox{${#1}$} \copy0 \kern-\wd0
      \kern.2pt \copy0 \kern-\wd0 \kern.2pt
    \copy0 \kern-\wd0 \kern.2pt \box0}
\def\pmxb#1{\setbox0\hbox{${#1}$} \copy0 \kern-\wd0
      \kern.2pt \copy0 \kern-\wd0 \kern.2pt
      \copy0 \kern-\wd0 \kern.2pt \copy0 \kern-\wd0 \kern.2pt \box0}
\def\pmxbb#1{\setbox0\hbox{${#1}$} \copy0 \kern-\wd0 \kern.2pt
      \copy0 \kern-\wd0 \kern.2pt
      \copy0 \kern-\wd0 \kern.2pt \copy0 \kern-\wd0 \kern.2pt
      \copy0 \kern-\wd0 \kern.2pt \box0}

\def\cdotss{\mathinner{\cdotp\cdotp\cdotp\cdotp\cdotp\cdotp\cdotp
        \cdotp\cdotp\cdotp\cdotp\cdotp\cdotp\cdotp\cdotp\cdotp\cdotp
        \cdotp\cdotp\cdotp\cdotp\cdotp\cdotp\cdotp\cdotp\cdotp\cdotp
        \cdotp\cdotp\cdotp\cdotp\cdotp\cdotp\cdotp\cdotp\cdotp\cdotp}}

\font\frak=eufm10 scaled\magstep1 \font\fak=eufm10 scaled\magstep2 \font\fk=eufm10
scaled\magstep3 \font\scriptfrak=eufm10 \font\tenfrak=eufm10


\mathchardef\za="710B  
\mathchardef\zb="710C  
\mathchardef\zg="710D  
\mathchardef\zd="710E  
\mathchardef\zve="710F 
\mathchardef\zz="7110  
\mathchardef\zh="7111  
\mathchardef\zvy="7112 
\mathchardef\zi="7113  
\mathchardef\zk="7114  
\mathchardef\zl="7115  
\mathchardef\zm="7116  
\mathchardef\zn="7117  
\mathchardef\zx="7118  
\mathchardef\zp="7119  
\mathchardef\zr="711A  
\mathchardef\zs="711B  
\mathchardef\zt="711C  
\mathchardef\zu="711D  
\mathchardef\zvf="711E 
\mathchardef\zq="711F  
\mathchardef\zc="7120  
\mathchardef\zw="7121  
\mathchardef\ze="7122  
\mathchardef\zy="7123  
\mathchardef\zf="7124  
\mathchardef\zvr="7125 
\mathchardef\zvs="7126 
\mathchardef\zf="7127  
\mathchardef\zG="7000  
\mathchardef\zD="7001  
\mathchardef\zY="7002  
\mathchardef\zL="7003  
\mathchardef\zX="7004  
\mathchardef\zP="7005  
\mathchardef\zS="7006  
\mathchardef\zU="7007  
\mathchardef\zF="7008  
\mathchardef\zW="700A  

\newcommand{\be}{\begin{equation}}
\newcommand{\ee}{\end{equation}}
\newcommand{\ra}{\rightarrow}
\newcommand{\lra}{\longrightarrow}
\newcommand{\bea}{\begin{eqnarray}}
\newcommand{\eea}{\end{eqnarray}}
\newcommand{\beas}{\begin{eqnarray*}}
\newcommand{\eeas}{\end{eqnarray*}}
\def\*{{\textstyle *}}
\newcommand{\R}{{\mathbb R}}
\newcommand{\T}{{\mathbb T}}
\newcommand{\C}{{\mathbb C}}
\newcommand{\I}{{\mathbb I}}
\newcommand{\unit}{{\mathbf 1}}
\newcommand{\SL}{SL(2,\C)}
\newcommand{\Sl}{sl(2,\C)}
\newcommand{\SU}{SU(2)}
\newcommand{\su}{su(2)}
\def\ssT{\sss T}
\newcommand{\G}{{\goth g}}
\newcommand{\D}{{\rm d}}
\newcommand{\Df}{{\rm d}^\zF}
\newcommand{\de}{\,{\stackrel{\rm def}{=}}\,}
\newcommand{\we}{\wedge}
\newcommand{\nn}{\nonumber}
\newcommand{\ot}{\otimes}
\newcommand{\s}{{\textstyle *}}
\newcommand{\ts}{T^\s}
\newcommand{\oX}{\stackrel{o}{X}}
\newcommand{\oD}{\stackrel{o}{D}}
\newcommand{\obD}{\stackrel{o}{\bD}}
\newcommand{\pa}{\partial}
\newcommand{\ti}{\times}
\newcommand{\A}{{\cal A}}
\newcommand{\Li}{{\cal L}}
\newcommand{\ka}{\mathbb{K}}
\newcommand{\find}{\mid}
\newcommand{\ad}{{\rm ad}}
\newcommand{\rS}{]^{SN}}
\newcommand{\rb}{\}_P}
\newcommand{\p}{{\sf P}}
\newcommand{\h}{{\sf H}}
\newcommand{\X}{{\cal X}}
\newcommand{\rB}{]_P}
\newcommand{\Ll}{{\pounds}}
\def\lna{\lbrack\! \lbrack}
\def\rna{\rbrack\! \rbrack}
\def\rnaf{\rbrack\! \rbrack_\zF}
\def\rnah{\rbrack\! \rbrack\,\hat{}}
\def\lbo{{\lbrack\!\!\lbrack}}
\def\rbo{{\rbrack\!\!\rbrack}}
\def\lan{\langle}
\def\ran{\rangle}
\def\zT{{\cal T}}
\def\tU{\tilde U}
\def\ati{{\stackrel{a}{\times}}}
\def\sti{{\stackrel{sv}{\times}}}
\def\aot{{\stackrel{a}{\ot}}}
\def\sati{{\stackrel{sa}{\times}}}
\def\saop{{\stackrel{sa}{\op}}}
\def\bwa{{\stackrel{a}{\bigwedge}}}
\def\svop{{\stackrel{sv}{\oplus}}}
\def\saot{{\stackrel{sa}{\otimes}}}
\def\cti{{\stackrel{cv}{\times}}}
\def\cop{{\stackrel{cv}{\oplus}}}
\def\dra{{\stackrel{\xd}{\ra}}}
\def\bdra{{\stackrel{\bd}{\ra}}}
\def\bAff{\mathbf{Aff}}
\def\Aff{\sss{Aff}}
\def\bHom{\mathbf{Hom}}
\def\Hom{\sss{Hom}}
\def\bt{{\boxtimes}}
\def\sot{{\stackrel{sa}{\ot}}}
\def\bp{{\boxplus}}
\def\op{\oplus}
\def\bwak{{\stackrel{a}{\bigwedge}\!{}^k}}
\def\aop{{\stackrel{a}{\oplus}}}
\def\ix{\operatorname{i}}
\def\V{{\cal V}}
\def\cD{{\cal D}}
\def\cC{{\cal C}}
\def\cE{{\cal E}}
\def\cL{{\cal L}}
\def\cN{{\cal N}}
\def\cR{{\cal R}}
\def\cJ{{\cal J}}
\def\cT{{\cal T}}
\def\cH{{\cal H}}
\def\cS{{\cal S}}
\def\cM{{\cal M}}
\def\bA{\mathbf{A}}
\def\bI{\mathbf{I}}
\def\wh{\widehat}
\def\wt{\widetilde}
\def\ol{\overline}
\def\ul{\underline}
\def\Sec{\sss{Sec}}
\def\Ph{\sss{Ph}}
\def\Lin{\sss{Lin}}
\def\ader{\sss{ADer}}
\def\ado{\sss{ADO^1}}
\def\adoo{\sss{ADO^0}}
\def\AS{\sss{AS}}
\def\bAS{\sss{AS}}
\def\bLS{\sss{LS}}
\def\bAP{\sss{AV}}
\def\bLP{\sss{LP}}
\def\AP{\sss{AP}}
\def\LP{\sss{LP}}
\def\LS{\sss{LS}}
\def\Z{\mathbf{Z}}
\def\oZ{\overline{\bZ}}
\def\oA{\overline{\bA}}
\def\cim{{C^\infty(M)}}
\def\de{{\cal D}^1}
\def\la{\langle}
\def\ran{\rangle}
\def\<{\langle}
\def\>{\rangle}
\def\bcS{\mathbb S}
\def\by{{\bi y}}
\def\bs{{\bi s}}
\def\bc{{\bi c}}
\def\bd{{\bi d}}
\def\bh{{\bi h}}
\def\bD{{\bi D}}
\def\bY{{\bi Y}}
\def\bX{{\bi X}}
\def\bL{{\bi L}}
\def\bV{{\bi V}}
\def\bW{{\bi W}}
\def\bS{{\bi S}}
\def\bT{{\bi T}}
\def\bC{{\bi C}}
\def\bE{{\bi E}}
\def\bF{{\bi F}}
\def\bP{{\bi P}}
\def\bH{{\bi H}}
\def\bp{{\bi p}}
\def\bz{{\bi z}}
\def\bZ{{\bi Z}}
\def\bq{{\bi q}}
\def\bQ{{\bi Q}}
\def\bx{{\bi x}}

\def\sA{{\sss A}}
\def\sC{{\sss C}}
\def\sD{{\sss D}}
\def\sG{{\sss G}}
\def\sH{{\sss H}}
\def\sI{{\sss I}}
\def\sJ{{\sss J}}
\def\sK{{\sss K}}
\def\sL{{\sss L}}
\def\sO{{\sss O}}
\def\sP{{\sss P}}
\def\sPh{{\sss P\sss h}}
\def\sT{{\sss T}}
\def\sV{{\sss V}}
\def\sR{{\sss R}}
\def\sS{{\sss S}}
\def\sE{{\sss E}}
\def\sF{{\sss F}}
\def\st{{\sss t}}
\def\sg{{\sss g}}
\def\sx{{\sss x}}
\def\sv{{\sss v}}
\def\sw{{\sss w}}
\def\sQ{{\sss Q}}
\def\sj{{\sss j}}
\def\sq{{\sss q}}
\def\xa{\tx{a}}
\def\xc{\tx{c}}
\def\xd{\tx{d}}
\def\xi{\tx{i}}
\def\xD{\tx{D}}
\def\xV{\tx{V}}
\def\xF{\tx{F}}


\setcounter{page}{1} \thispagestyle{empty}
\bigskip

\bigskip

\title{The Schr\"odinger operator as a generalized Laplacian}

        \author{
        Katarzyna  Grabowska$^1$, Janusz Grabowski$^2$, Pawe\l\ Urba\'nski$^1$\\
        \\
         $^1$ {\it Physics Department}\\
                {\it University of Warsaw} \\
         $^2$ {\it Institute of Mathematics}\\
                {\it Polish Academy of Sciences}
                }
\date{}
\maketitle
\begin{abstract}
The Schr\"odinger operators on the Newtonian space-time are defined in a way which make
them independent on the class of inertial observers. In this picture the Schr\"odinger
operators act not on functions on the space-time but on sections of certain
one-dimensional complex vector bundle -- the {\em Schr\"odinger line bundle}. This line
bundle has trivializations indexed by inertial observers and is associated with an
$U(1)$-principal bundle with an analogous list of trivializations -- the {\em
Schr\"odinger principal bundle}. If an inertial frame is fixed, the Schr\"odinger
bundle can be identified with the trivial bundle over space-time, but as there is no
canonical trivialization (inertial frame), these sections interpreted as
`wave-functions' cannot be viewed as actual functions on the space-time. In this
approach the change of an observer results not only in the change of actual coordinates
in the space-time but also in a change of the phase of wave functions. For the
Schr\"odinger principal bundle a natural differential calculus for `wave forms' is
developed that leads to a natural generalization of the concept of Laplace-Beltrami
operator associated with a pseudo-Riemannian metric. The free Schr\"odinger operator
turns out to be the Laplace-Beltrami operator associated with a naturally distinguished
invariant pseudo-Riemannian metric on the Schr\"odinger principal bundle. The presented
framework does not involve any {\em ad hoc} or axiomatically introduced geometrical
structures. It is based on the traditional understanding of the Schr\"odinger operator
in a given reference frame -- which is supported by producing right physics predictions
-- and it is proven to be strictly related to the frame-independent formulation of
analytical Newtonian mechanics and Hamilton-Jacobi equations, that makes a bridge
between the classical and quantum theory.


\end{abstract}
\section{Introduction}
In the papers \cite{GGU1,GGU2,GGU3,U} we have presented an approach to differential geometry in which sections
of a one-dimensional affine bundle over a manifold have been used instead of functions on the manifold. This
approach, initiated by W.~M.~Tulczyjew in \cite{TU,TUZ}, has been successfully applied to frame-independent
description of different systems, in particular to a frame-independent formulation of Newtonian mechanics
\cite{GU}.

The latter problem is closely related to the problem of frame-independent formulation of
wave mechanics in the Newtonian space-time. It is known that a solution of the
Schr\"odinger equation in one inertial frame will not, in general, satisfy the
Schr\"odinger equation in a different frame. The same quantum state of a particle must be
represented by a different wave function in reference to a different inertial frame. The
corresponding gauge transformation of solutions of the Schr\"odinger equation was known
already to W.~Pauli \cite{Pa}. Many ways of solving this problem have been proposed in
the literature. For instance, a general axiomatic theory of quantum bundles, quantum
metrics, quantum connections etc. has been developed in \cite{JM} to deal with a
covariant description of Schr\"odinger operators in curved space-times. Another general
fibre bundle formulation of nonrelativistic quantum mechanics has been proposed in a
series of papers \cite{Il}.

An approach which is the closest to what we propose in this paper is a frame-independent formulation of wave
mechanics by extending the Newtonian space-time to five-dimensional Galilei space \cite{DB,WMT2,WMT3}. The
corresponding geometry is associated with the Bargmann group - nontrivially extended Galilei group \cite{Ba}.

\medskip
In the present paper we change this view-point a little bit, making the `wave
functions' living on a four-dimensional base again. For simplicity, we deal with the
flat Newtonian space-time and the very standard Schr\"odinger operators to show that a
frame-independent formulation of wave mechanics (for every mass $m\ne 0$) is possible
in terms of a principal $U(1)$-bundle $P_m$ -- the {\em Schr\"odinger principal
bundle}. For a fixed inertial frame this bundle can be identified with the trivial
bundle over the space-time, but no canonical trivialization is given. With this bundle
there is associated a complex line bundle $L_m$ -- the {\em Schr\"odinger line bundle}.
Only the projective class of this bundle is uniquely defined, which is associated with the fact
that wave functions are
sometimes understood as defined up to a phase factor. In our picture, the Schr\"odinger
operator acts not on functions on the space-time but on sections of $L_m$. This bundle,
constructed from the data provided by all possible inertial observers, has no canonical
trivialization, so its sections cannot be viewed as functions on the space-time.
Indeed, they change under the change of an inertial frame in a way which is different
from the way functions do. We would like to stress that this point causes often
difficulties for some people who have problems with distinguishing trivializable
bundles from trivial ones. This distinction should be taken seriously while reading
this paper. One can simply explain this problem in plain English by pointing out that
`mortal' is not the same as `dead'. One can interpret this fact in the way that passing
to another observer leads not only to a certain change in positions and velocities but
also to a change in the phase of wave functions.

\medskip
Having constructed the Schr\"odinger principal bundle $P_m$ as the proper geometrical tool for understanding
the Schr\"odinger operators, we develop a differential calculus based on the {\em Atiyah Lie algebroid} $\A_m$
associated with this bundle and applied for {\it wave forms} being sections of $\bigwedge^k\A^*_m\ot L_m$.
Mathematically it is a version of the deformation of the de Rham differential considered by E.~Witten
\cite{Wi} and similar to the calculus for {\em Jacobi algebroids} as developed in \cite{IM, GM1,GM2}. With
this calculus, gradients and divergences, so (generalized) Laplace-Beltrami operators, associated with
pseudo-Riemannian metrics are naturally defined. This construction, applied to a naturally distinguished
pseudo-Riemannian metric on $P_m$, allows us to write the free Schr\"odinger operator
$$\cS^0_m\zc=\frac{\hbar^2}{2m}\sum_k\frac{\pa^2\zc}{\pa{y_k}^2}+i\hbar\frac{\pa\zc}{\pa_t}$$
as proportional to the corresponding Laplace-Beltrami operator.

\medskip
We want to stress three facts. First, we do not look just for transformations rules for
solutions of the Schr\"odinger equation in different reference frames, but we build a
bundle, sections of which represent the arguments of the Schr\"odinger operator (`wave
functions') that gives to the operator itself a covariant geometrical meaning.
Moreover, we show that the projective class of transformation rules, so the projective
class of the Schr\"odinger bundle, is unique. All known to us constructions of this
type are based on explicit or hidden assumptions concerning the dynamics of a Newtonian
particle. For example, assumptions that an intrinsic Lagrangian is a function on the
time-configuration-velocity space, or that the energy-momentum phase space is the
cotangent bundle of the Newtonian space-time. On the other hand, it became clear
nowadays that an intrinsic, i.e., a frame-independent formulation of the Newtonian
dynamics requires affine and not vectorial objects. We refer here to our earlier work
\cite{GGU1,GGU2,GU,U}, to recent papers by Jany\v ska and Modugno \cite{JM}, and
Mangiarotti and Sardanashvily \cite{MS}.

Second, we are able to interpret the standard Schr\"odinger operator as a (generalized) Laplace-Beltrami
operator. To do that one has to use a deformed differential calculus, based on a de Rham-like differential
which is similar to the one considered by E.~Witten \cite{Wi} and to the differential in the theory of so
called {\em Jacobi algebroids} \cite{IM, GM1, GM2}. In this calculus, the Laplace-Beltrami operator associated
with a naturally distinguished invariant pseudo-Riemannian metric on the Schr\"odinger principal bundle turns
out to coincide up to a factor with the classical free (with the potential 0) Schr\"odinger operator. In our
opinion, this idea may find much broader applications than just the ones present in our paper.

And last but not least, we prove that the proposed formulation is strictly related to
the frame-independent formulation of analytical Newtonian mechanics \cite{GU}. The
"logarithm" $\Z_m$ of the principal Schr\"odinger bundle is namely an $\R$-principal
bundle, so an {\it affine values bundle (AV-bundle)} in the terminology of
\cite{GGU1,GGU2,GGU3,GU,U}. The Hamiltonian bundle, i.e. an AV-bundle whose sections
represent possible Hamiltonians, constructed out of it coincides with the bundle
obtained in \cite{GGU2,GU} for the Newtonian particle with mass $m$. This means that
the bundle $\Z_m$ is a Hamilton-Jacobi bundle for the Newtonian particle with mass $m$,
i.e. it is an AV-bundle whose sections are subject of the affine (frame-independent)
Hamilton-Jacobi equations. This makes a bridge between the classical and quantum theory
which, in our opinion, is not understood completely yet and almost not present in the
literature. The nice relation of the constructed Schr\"odinger bundle to the intrinsic
Lagrangian or Hamiltonian bundle of a massive Newtonian particle we view as an evidence
that our description is proper. In this sense, the present work is a natural step
following the series of papers \cite{GGU1}-\cite{GGU3} in which we have developed the
geometry of affine values and applied it to frame-independent formulation of Classical
Mechanics.

The paper is organized as follows. We start with recalling the Newtonian picture for the space-time and the
standard Schr\"odinger operators associated with potentials on it. Then, we present the main idea of what a
`wave function' and the Schr\"odinger operator should be and, in Section 3, we present the idea of a principal
or vector bundle with a distinguished set of trivializations.

In section 4 we find the unique form of the transformation rules in the trivial complex line bundle over
$\R^3\ti\R$ that leave the Schr\"odinger operators invariant. These transformations rules are used in
constructing the Schr\"odinger principal $U(1)$-bundle $P_m$ and the Schr\"odinger line bundle $L_m$ (for
fixed `mass' $m$). The `wave functions' are understood as sections of $L_m$ and, for every fixed potential
$U$, the Schr\"odinger operator $\bcS^U_m$ associated with this potential is a well-defined second-order
differential operator on $L_m$. This description is our frame-independent interpretation of the Schr\"odinger
operators.

In Section 5 we show that the above description agrees with the frame-independent
description of the Newtonian mechanics and that there is a close relation of the
Schr\"odinger bundles with the affine bundles whose sections are interpreted as subject
of the Hamilton-Jacobi equations and whose phase bundle gives rise to an affine
Hamiltonian formalism, as defined in \cite{GGU2,GU}.

A differential calculus for wave forms, i.e. sections of the bundles
$\left(\bigwedge^k\A_m^\ast\right)\ot_N F_m$, where $\A_m^\ast$ is the bundle dual to
the so called {\it Atiyah Lie algebroid} $\A_m$ associated with the principal bundle
$P_m$, is developed in Section 6.

Section 7 is devoted to finding a naturally distinguished pseudo-Riemannian metric $\zm_m$ on $P_m$ -- the
{\it Schr\"odinger metric} -- which, in coordinates associated with any inertial frame, extends the standard
spatial Euclidean metrics in the space-time and which looks exactly in the same way for all inertial
observers. We find also the volume form associated with this metric.

The above-mentioned metric and the volume are used in the next section to define the corresponding `gradient
wave-vector fields', associated with `wave functions', and wave-divergences associates with the gradients, so,
in turn, the corresponding (generalized) Laplace-Beltrami operator. This operator actc on wave-functions and
coincides, up to a factor, with the free Schr\"odinger operator we started with.

\section{Newtonian space-time}

The {\it Newtonian space-time} (some authors prefer to call it {\em Galilean space-time},
but we follow the terminology of Benenti \cite{Be} and Tulczyjew \cite{WMT2}) is a system
$(N,\tau, g)$, where $N$ is a four-dimensio\-nal affine space for which, say $V$, is the
model vector space, where $\tau$ is a non-zero element of $V^\ast$, and where $g\colon
E_0\rightarrow E_0^\ast$ represents an Euclidean metric on $E_0=\ker\tau$. The
corresponding scalar product reads $\< v\mid v'\>=(g(v))(v')$ and the corresponding norm
$\Vert v\Vert=\sqrt{\< v\mid v\>}$. The elements of the space $N$ represent events. The
time elapsed between two events is measured by $\tau$:
    $$\Delta t(x,x')=\tau(x-x')$$
and the distance between two simultaneous events is measured by $g$:
    $$d(x,x')=\Vert x-x'\Vert.$$
    The space-time $N$ is fibred over the time $\T=N\slash E_0 $
which is a one-dimensional affine space modelled on $\R$.

Let $E_1$ be an affine subspace of $V$ defined by the equation
$\tau(v)=1$. The model vector space for this subspace is $E_0$. An
element of $E_1$ represents velocity of a particle. The affine
structure of $N$ allows us to associate to an element $u$ of $E_1$
the family of inertial observers that move in the space-time with
the constant velocity $u$. In this way we can interpret an element
of $E_1$ also as a {\it class of inertial reference frames} while
an {\it inertial reference frame} is understood as a pair
$(x_0,u)\in N\times E_1$. For a fixed inertial frame $(x_0,u)$, we
can identify $N$ with $E_0\ti\R$ by \be \zF_{(x_0,u)}:N\ra
E_0\ti\R,\quad x\mapsto \left((x-x_0)-
\zt(x-x_0)u,\zt(x-x_0)\right).\ee A change of the inertial
reference frame results in the change of this identification and
it is represented by
\bea\label{1}\zY^{(x_0',u')}_{(x_0,u)}&=&\zF_{(x'_0,u')}\circ\zF_{(x_0,u)}^{-1}:
E_0\ti\R\ra E_0\ti\R, \\ (v,t)&\mapsto &
(v-((x'_0-x_0)-\zt(x'_0-x_0)u')-(u'-u)t,t-\zt(x'_0-x_0)).\eea We can fix orthonormal
linear coordinates $y=(y_i):E_0\ra\R^3$ in $E_0$ so that $\Vert
v\Vert^2=\sum_iy_i^2(v)$. Then, with every inertial frame $(x_0,u)$, we can associate
coordinates $(y,t)$ in $N$, thus $V$, with
$(y,t)(x)=\zf_{(x_0,u)}(x)=(y(x-x_0-\zt(x-x_0)u),\zt(x-x_0))$, and the change of
coordinates $\zvy^{(x_0',u')}_{(x_0,u)}$ corresponding to $\zY^{(x_0',u')}_{(x_0,u)}$
reads
\be\label{2}\zvy^{(x_0',u')}_{(x_0,u)}(y,t)=\zf_{(x'_0,u')}\circ\zf_{(x_0,u)}^{-1}(y,t)=
\left(y+w_u+y(v)(t+t_0),\,t+ t_0\right), \ee where
$(w_u,t_0)=(y(x_0-x_0'-\zt(x_0-x_0')u),\zt(x_0-x_0'))\in\R^3\ti\R$
are coordinates of $w=x_0-x_0'\in V$ for the observer $(x_0,u)$
and $y(v)\in\R^3$ are coordinates of $v=u-u'\in E_0$. Note that
the maps $\zvy^{(x_0',u')}_{(x_0,u)}$ are affine transformations
that satisfy the {\it cocycle condition}
$\zvy^{(x_0'',u'')}_{(x_0',u')}\circ \zvy^{(x_0',u')}_{(x_0,u)}
=\zvy^{(x_0'',u'')}_{(x_0,u)}$. Thus, we have
\bea\label{2a}\zvy^{(x_0'',u'')}_{(x_0',u')}(y,t)&=&\zvy^{(x_0'',u'')}_{(x_0,u)}\circ
\left(\zvy^{(x_0',u')}_{(x_0,u)}\right)^{-1}=\left(y+w_{u'}+y(v')(t+t_0'),t+t_0'\right)\\
&=&(y+w_u'+y(v')(t+t_0')+y(v)t_0',\,t+ t_0')\,, \nn
\eea
where $(w_u',t_0')$ are coordinates of $w'=x_0'-x_0''$ for the
observer $(x_0,u)$ and $v'=u'-u''$.
\section{The Schr\"odinger operator and principal bundles with  trivializations}
The classical Schr\"odinger operator in coordinates $(y,t)\in\R^3\ti\R$, for a particle of mass $m$ and a
potential $\wt{U}\in C^\infty(\R^3\ti\R)$, is a second order complex differential operator which reads
\be\label{S} \cS^{\wt{U}}_m\zc=\frac{\hbar^2}{2m}\sum_k\frac{\pa^2\zc}{\pa y_k^2}+i\hbar\frac{\pa\zc}{\pa
t}-\wt{U}\zc. \ee Here, $\sum_k\frac{\pa^2}{\pa y_i^2}$ is clearly the spatial Laplace-Beltrami operator
associated with the metric $g$. The problem is that, if assumed as acting on functions, the Schr\"odinger
operator (\ref{S}) is not invariant with respect to the change of coordinates (\ref{2}) associated with the
choice of another inertial frame. On the other hand, by  arguments coming from physics, the form of the
Schr\"odinger operator should be independent on the choice of an inertial observer.

The solution we propose is that the Schr\"odinger operator acts in
fact on sections of certain 1-dimensional complex vector bundle
$L_m$ over $N$ (we will call it {\em Schr\"odinger bundle}) which
is trivializable (with a list of distinguished trivializations)
but with no canonical trivialization. A change of an observer
results not only in a change of coordinates  but also in the
change of the phase of the wave function. Thus the situation is
parallel to the one we encounter in frame-independent description
of the standard lagrangian in Newtonian mechanics \cite{GU}.

To be more precise, let us recall that a principal or a vector
bundle is defined by an atlas of local identifications of our
structure with the trivial ones such that the transition maps
respect the structure. One assumes often {\it a priori} that the
atlas is maximal. Here, however, we will understand the given and
not maximal atlas as an immanent part of the structure. This is
because we want the transition maps to preserve an additional
structure. This means that a {\it $U(1)$-bundle $P$ with
trivializations} is understood as a smooth manifold equipped with
a family of (global) trivializations $\Psi_\zl:P\ra M\ti U(1)$,
$\zl\in\zL$, over a manifold $M$ such that the transition maps
$$T^{\zl'}_\zl=\Psi_{\zl'}\circ\Psi_\zl^{-1}:M\ti U(1)\ra M\ti U(1)$$ are $U(1)$-bundle
isomorphisms, i.e. they are of the form
\be\label{tryw}T^{\zl'}_\zl(x,z)=\left(\zvy^{\zl'}_\zl(x),
e^{i\wt{F}^{\zl'}_\zl\circ\zvy^{\zl'}_\zl(x)}\cdot z\right)\,,
\ee
where $\wt{F}^{\zl'}_\zl:M\ra\R$ are smooth functions. As a consequence, $P$ carries a
unique structure of a principal $U(1)$-bundle over $M_0=P/U(1)$ and the family
$(\Psi_\zl)_{\zl\in\zL}$ of distinguished trivializations over $M$ defines a family of
distinguished sections $(\zc_\zl)_{\zl\in\zL}$, where $\zc_\zl:M_0\ra P$ is defined by
$\Psi_\zl\left(\zc_\zl(M_0)\right)=M\ti\{ 1\}$. Note that the {\it pull-back} of a
section $\zc$ of the trivial principal bundle $M\ti U(1)$, induced by the transition
map $T^{\zl'}_{\zl}$ reads
\be\label{pullback}
(T^{\zl'}_{\zl})^\ast\zc=\left(e^{-i\wt{F}^{\zl'}_\zl}\cdot\zc\right)\circ\zvy^{\zl'}_\zl\,.
\ee
However, as the pull-back reverses the order of composition
$$(T^{\zl''}_{\zl'}\circ
T^{\zl'}_{\zl})^\ast=(T^{\zl'}_{\zl})^\ast\circ(T^{\zl''}_{\zl'})^\ast\,,
$$
we will prefer to use the {\it push-forwards},
$\left(T^{\zl'}_{\zl}\right)_\ast=\left(\left(T^{\zl'}_{\zl}\right)^\ast\right)^{-1}$,
\be\label{sectransf}
(T^{\zl'}_{\zl})_\ast\zc=\left(e^{i\wt{F}^{\zl'}_\zl}\cdot\zc\right)\circ(\zvy^{\zl'}_\zl)^{-1}
\ee
instead. It is also clear that multiplying every trivialization
$\Psi_\zl$ of the $U(1)$-bundle $P$ by a complex number $z_\zl\in
U(1)$ (a phase) will give data for another family of
trivializations. More precisely, for $z_0\in U(1)$ denote by
$\wh{z}_0$ the action of $z_0$ on the principal $U(1)$ bundle
$M\ti U(1)$. Then, for any map $\zL\ni\zl\mapsto z_\zl\in U(1)$,
$\wh{\Psi}_\zl=\wh{z}_\zl\circ\Psi_\zL$ is another list of
trivializations of $P$. We will say that these principal
$U(1)$-bundles with trivializations are {\it in the same
projective class} $[P]$. The same can be repeated for {\it complex
line bundles} with trivializations constructed out of these
trivializations of principal $U(1)$-bundles, i.e. for the
corresponding associated complex line bundles.

\medskip
Let us stress the fact that isomorphism of such structures depend on an identification of two distinguished
atlases, so that principal bundles with trivializations may be not isomorphic {\bf as bundles with
trivializations} even being isomorphic as principal bundles.
\begin{definition}{\rm By {\it principal $U(1)$-bundle with trivializations over a manifold $M$}
we understand a manifold $P$ together with a map $\Psi:\zL\ra
Diff(P, M\ti U(1))$ from a set $\zL$ to the set of diffeomorphisms
$\zf:P\ra M\ti U(1)$, $\zl\mapsto\Psi_\zl$, such that the {\it
transition maps} $T^{\zl'}_\zl=\Psi_{\zl'}\circ\Psi_\zl^{-1}:M\ti
U(1)\ra M\ti U(1)$ respect the $U(1)$-bundle structure,
$$T^{\zl'}_\zl(x,z)=\left(\zvy^{\zl'}_\zl(x),e^{i\wt{F}^{\zl'}_\zl\circ\zvy^{\zl'}_\zl(x)}\cdot z\right)\,,
$$
so that they define a principal $U(1)$-bundle structure on $P$. A {\it projective
morphism} of principal $U(1)$-bundles  with trivializations $(P,\Psi)$ and
$(\wt{P},\wt{\Psi})$ consists of a map $j:\zL\ra\wt{\zL}$ and a $U(1)$-bundle morphism
$J:P\ra\wt{P}$ such that, for each $\zl\in\zL$,
$$\Xi_\zl=\wt{\Psi}_{j(\zl)}\circ J\circ(\Psi_\zl)^{-1}:M\ti U(1)\ra \wt{M}\ti U(1)$$
is a morphism of principal $U(1)$ bundles which is of the form
$$\Xi_\zl(x,z)=(\phi_\zl(x),z_\zl\cdot z)\,,
$$
i.e. which is constant on $U(1)$ up to a multiplication by the constant $z_\zl\in
U(1)$. A projective morphism we call {\it morphism} if the constants are trivial,
$z_\zl=1$, i.e., if $\Xi_\zl$ is identity on $U(1)$.

A projective morphism as above is a {\it projective isomorphism}
if the map $j$ is bijective and $J$ is an isomorphism of principal
bundles . A {\it projective class} $[P]$ of a principal
$U(1)$-bundle with trivializations consists of all principal
$U(1)$-bundles with trivializations that are projectively
isomorphic to $(P,\Psi)$. Again, for {\it isomorphisms of
principal $U(1)$-bundles with trivializations}, the map $j$ is
bijective and $J$ is an isomorphism of principal bundles.
 }
\end{definition}

\bigskip\noindent In the above sense, a {\it trivial bundle} is a
bundle with just one trivialization and it is not isomorphic with
bundles with the set of trivializations containing more than one
element, since there is no way to distinguish one trivialization
from another. Moreover, isomorphisms between trivial bundles can
be identified with diffeomorphisms between base manifolds. More
precisely, they are of the form $\wh{\phi}:M\ti U(1)\ra\wt{M}\ti
U(1)$, $\wh{\phi}(x,z)=(\phi(x),z)$, where $\phi:M\ra\wt{M}$ is a
diffeomorphism.
\begin{theorem} $U(1)$-principal bundles with trivializations $(P,\Psi)$ and $(\wt{P},\wt{\Psi})$ are isomorphic
(resp., projectively isomorphic) if and only if there is a bijection $j:\zL\ra\wt{\zL}$ and a $U(1)$-bundle
isomorphism $J:P\ra \wt{P}$  that relates (relates, up to a constant factor) the distinguished sections
$\zc_\zl$ and $\wt{\zc}_{j(\zl)}$ for all $\zl\in\zL$.
\end{theorem}
\proof Suppose a bijection $j:\zL\ra\wt{\zL}$ and a $U(1)$-bundle isomorphism $J:P\ra\wt{P}$ define an
isomorphism. Since $J\circ\left(\Psi_\zl\right)^{-1}=\left(\wt{\Psi}_{j(\zl)}\right)^{-1}\circ\Xi_\zl$,
\beas J(\zc_\zl(M_0))&=&J\left(\left(\Psi_\zl\right)^{-1}\left(M\ti\{ 1\}\right)\right)=
\left(\wt{\Psi}_{j(\zl)}\right)^{-1}\left(\Xi_\zl\left(M\ti\{ 1\}\right)\right)\\
&=&\left(\wt{\Psi}_{j(\zl)}\right)^{-1}\left(\wt{M}\ti\{
1\}\right)=\wt{\zc}_{j(\zl)}\left(\wt{M}_0\right)\,,\eeas so the sections ${\zc}_{\zl}$ and
$\wt{\zc}_{j(\zl)}$ are $J$-related.

\medskip\noindent
Conversely, if ${\zc}_{\zl}$ and $\wt{\zc}_{j(\zl)}$ are $J$-related, then
$$J\left(\left(\Psi_\zl\right)^{-1}\left(M\ti\{ 1\}\right)\right)=\left(\wt{\Psi}_{j(\zl)}\right)^{-1}\left(\wt{M}\ti\{
1\}\right)\,,$$ so that
$$\Xi_\zl=\wt{\Psi}_{j(\zl)}\circ J\circ(\Psi_\zl)^{-1}(x,z)=(\phi_\zl(x),z)$$
and the trivializations are isomorphic.

\medskip\noindent
The proof in the projective case is analogous.
\endproof

\medskip\noindent The transition maps satisfy automatically the {\it cocycle condition}
\be\label{cc1a}T^{\zl}_\zl=id\,,\quad
 T^{\zl''}_{\zl'}\circ T^{\zl'}_\zl=T^{\zl''}_\zl
\ee
which can be rewritten in the form
\be\label{cc2}\zvy^{\zl}_\zl=id\,,\quad \zvy^{\zl''}_{\zl'}\circ\zvy^{\zl'}_\zl=\zvy^{\zl''}_{\zl},\quad
\wt{F}^{\zl}_\zl=0\,,\quad
\wt{F}^{\zl''}_{\zl}\circ\zvy^{\zl''}_{\zl'}=\wt{F}^{\zl''}_{\zl'}\circ\zvy^{\zl''}_{\zl'}+\wt{F}^{\zl'}_\zl\,.
\ee
The cocycle condition can be interpreted as the fact that
$$T:\zL\ti\zL\ni(\zl',\zl)\mapsto T^{\zl'}_\zl\in Aut(M\ti U(1))
$$
is a morphism of the pair groupoid $\zL\ti\zL$
into the group $Aut(M\ti U(1))$ of automorphisms of the principal bundle $M\ti U(1)$.

Of course, as easily seen, one can start with transition maps
(\ref{tryw}) satisfying the cocycle condition (\ref{cc2}) and
construct the corresponding principal bundle with trivializations
up to isomorphism by taking $P$ to be the space of classes in
$\zL\ti M\ti U(1)$ with respect to the equivalence relation
\be\label{construction}[\zl,x,z]\sim[\zl',x',z']\Leftrightarrow T^{\zl'}_\zl(x,z)=
\left(\zvy^{\zl'}_\zl(x),\,e^{i\wt{F}^{\zl'}_\zl\circ\zvy^{\zl'}_\zl(x)}\cdot
z\right)=(x',z')\,.
\ee
This is canonically a principal $U(1)$-bundle with respect to the action
$z_0[\zl,x,z]=[\zl,x,z_0\cdot z]$ with a family $\Psi_\zl$ of trivializations indexed
by $\zL$ and defined by
$$\Psi_\zl([\zl,x,z])=(x,z)\in M\ti U(1)\,.$$
The transition functions for these trivializations coincide with $T^{\zl'}_\zl$.

It is completely obvious that the data given by transition maps for a principal $U(1)$
bundle can be used to construct a unique (up to isomorphism) complex line bundle. Our
{\it Schr\"odinger complex line bundle} $L_m$ (for the mass $m$) will be obtained as a
complex vector bundle with the model fibre $\C$ -- associated with a principal
$U(1)$-bundle $P_m$ with trivializations indexed by inertial observers -- the {\it
Schr\"odinger principal bundle}. Since one often regards wave functions as being
defined up to a constant phase, it is only the projective class of a $U(1)$-bundle with
trivializations that really matters. We will see that all possible Schr\"odinger
bundles are in the same class which means uniqueness of this structure.

Similarly like a principal $U(1)$-bundle with trivializations can be defined up to isomorphism by the family
of transition functions $T^{\zl'}_{\zl}$ satisfying the cocycle conditions (\ref{cc1a}), the projective class
of a principal $U(1)$-bundle with  trivializations can be defined up to projective isomorphism by the family
of transition functions $T^{\zl'}_{\zl}$ satisfying the cocycle conditions (\ref{cc1a}) up to constants (we
will call such $T$ a {\it projective cocycle}):
\be\label{projclass}T^{\zl}_\zl(x,z)=\wh{z}_{\zl}\,,\quad T^{\zl''}_{\zl'}\circ
T^{\zl'}_\zl\circ\left(T^{\zl''}_\zl\right)^{-1}(x,z)=\wh{z}_{(\zl'',\zl',\zl)}\,.
\ee
Indeed, let us choose $\zl_0$ and define a new family of `transition functions'
$\wt{\Psi}_\zl=T^{\zl}_{\zl_0}$,
$$\wt{T}^{\zl'}_\zl=T^{\zl'}_{\zl_0}\circ\left(T^\zl_{\zl_0}\right)^{-1}\,.$$
Then $\wt{T}^\zl_\zl=id$ and $\wt{T}^{\zl''}_{\zl'}\circ \wt{T}^{\zl'}_\zl=\wt{T}^{\zl''}_\zl$, so the family
$\wt{T}^{\zl'}_\zl$ satisfies the cocycle condition and gives rise to a well-defined principal $U(1)$ bundle
with trivializations. If we choose in the above construction another $\zl_0$, say $\zl_1$, then the family of
transition maps
$$T^{\zl'}_{\zl_1}\circ\left(T^\zl_{\zl_1}\right)^{-1}$$
differs from $\wt{T}^{\zl'}_\zl$ by constant factors, so defines a principal $U(1)$
bundle with trivializations in the same projective class,
\begin{theorem}\label{uniq} A map $T:\zL\ti\zL\ra Aut(M\ti U(1))$, $(\zl',\zl)\mapsto
T^{\zl'}_\zl$, satisfying the cocycle condition (\ref{cc1a}) (resp., the cocycle condition up to constants
(\ref{projclass})), defines canonically a principal $U(1)$-bundle with trivializations indexed by $\zL$ up to
isomorphism (resp., up to projective isomorphism).
\end{theorem}

\section{The Schr\"odinger bundles}
The Schr\"odinger complex line bundle will have trivializations enumerated by inertial
observers $\zl=(x_0,u)$. We have to combine every change of coordinates (\ref{2}) in
$N$ with a linear change in values of wave functions
\be\label{ccx}T^{(x_0'',u'')}_{(x_0',u')}(y,t,z)=
\left(\zvy^{(x_0'',u'')}_{(x_0',u')}(y,t),\,e^{F^{(x_0'',u'')}_{(x_0',u')}\circ
\zvy^{(x_0'',u'')}_{(x_0',u')}(y,t)}\cdot z\right)\,,\ee so that the push-forward of
wave-functions
\be\label{cc}\left(T^{(x_0'',u'')}_{(x_0',u')}\right)_\ast(\zc)(y,t)=e^{F^{(x_0'',u'')}_{(x_0',u')}(y,t)}
\cdot\zc\left(\left(\zvy^{(x_0'',u'')}_{(x_0',u')}\right)^{-1}(y,t)\right)
\ee
preserves the form of the Schr\"odinger operator. Of course, as mentioned above, there
is an obvious freedom in constructing such a line bundle, as we can always put
$$\wt{F}^{(x_0'',u'')}_{(x_0',u')}=F^{(x_0'',u'')}_{(x_0',u')}+A(x_0'',u'')-A(x_0',u')$$
for any function $A:N\ti E_1\ra\C$, as the cocycle condition is
automatically satisfied and the multiplication by a constant
function commutes with the Schr\"odinger operator. We will see
later on that this is the only freedom admitted by our conditions.

At the beginning we can simplify this problem a little bit. Since,
as can be easily seen, the part corresponding to the potential
$\wt{U}$ associated with a function $U$ on $N$ behaves properly
and the Schr\"odinger operator is invariant with respect to the
change of coordinates associated with observers moving with the
same velocity, $u=u'$, we can assume that $\wt{U}=0$ and
$x'_0=x_0$. Thus we shall look for an action of the commutative
group $E_0$ in $\R^3\ti\R\ti\C$ of the form
\be\label{4}R_v(y,t,z)=
\left(y+y(v)t,t,e^{F_v(y+y(v)t,t)}z\right),\ee corresponding to
the representation of $E_0$ in the algebra
$C^\infty_\C(\R^3\ti\R)$ of complex-valued functions on
$\R^3\ti\R$,
\be\label{4a}(R_v)_\ast(\zc)(y,t)=e^{F_v(y,t)}\zc(y-y(v)t,t),\ee such
that the "free" Schr\"odinger operator
\be\label{S1}\cS^0_m\zc=\frac{\hbar^2}{2m}\sum_k\frac{\pa^2\zc}{\pa
y_i^2}+i\hbar\frac{\pa\zc}{\pa t}\ee remains unchanged:
\be\label{5} \cS^0_m\left(e^{F_v(y,t)}\zc(y-y(v)t,t)\right)=
e^{F_v(y,t)}\cS^0_m(\zc)(y-y(v)t,t).\ee

\begin{remark} That our spatial part is 3-dimensional is motivated by physics. However,
from the mathematical point of view, there is no difference if we use other dimensions.
All considerations and proofs remain unchanged if we use $\R^n\ti\R\ti\C$ instead of
$\R^3\ti\R\ti\C$.
\end{remark}

Let us look what the function $F_v$ should be, in order that
(\ref{5}) is satisfied. Straightforward calculations, where we put
for simplicity $y(v)=v=(v_k)$, show that (\ref{5}) is equivalent
to
\bea &\zc(y-vt,t)\left(i(\pa_tF_v)(y,t)+\frac{\hbar}{2m}\left(
\sum_k(\pa_{y_k}F_v)^2(y,t)+\sum_k(\pa_{y_k}^2F_v)(y,t)\right)\right)+\\
& \sum_k(\pa_{y_k}\zc)(y-vt,t)\left(\frac{\hbar}{m} (\pa_{y_k}F_v)(y,t)-iv_k\right)=0\nn\eea for all complex
functions $\zc$ on $\R^3\ti\R$. Since $\zc$ is arbitrary, this, in turn, is equivalent to the system of
equations \bea\label{r1} &i(\pa_tF_v)(y,t)+\frac{\hbar}{2m}\left(\sum_k(\pa_{y_k}F_v)^2(y,t)+
\sum_k(\pa_{y_k}^2F_v)(y,t)\right)=0,\\
&\frac{\hbar}{m}(\pa_{y_k}F_v)(y,t)-iv_k=0\,,k=1,2,3\,.\label{r2}
\eea From (\ref{r2}) it follows that $\pa_{y_k}^2F_v=0$, $k=1,2,3$,
so that (\ref{r1}) reduces to
\be\label{r1a}i(\pa_tF_v)(y,t)-\frac{m}{2\hbar}\sum_kv_k^2=0. \ee
The equations (\ref{r2}) and (\ref{r1a}) for partial derivatives determine $F_v$ up to a constant, so, as can
be easily seen,
\be\label{F}F_v(y,t)=\frac{im}{\hbar}\left(
\sum_kv_ky_k-\frac{t}{2}\sum_kv_k^2\right)+c\,. \ee  Going back to the general case we
conclude that the transformation rule (\ref{cc}) that preserves the form of the
Schr\"odinger operator requires that
$$F^{(x_0'',u'')}_{(x_0',u')}(y,t)=\frac{im}{\hbar}\left(
\sum_kv_k'y_k-\frac{t}{2}\sum_k(v_k')^2\right)+c^{(x_0'',u'')}_{(x_0',u')}\,.
$$
The cocycle condition
\be\label{cc1}T^{(x_0'',u'')}_{(x_0',u')}\circ
T^{(x_0',u')}_{(x_0,u)}=T^{(x_0'',u'')}_{(x_0,u)}\ee yields now
that
$$F^{(x_0'',u'')}_{(x_0',u')}=F^{(x_0'',u'')}_{(x_0,u)}-F^{(x_0',u')}_{(x_0,u)}\circ\left(
\zvy^{(x_0'',u'')}_{(x_0',u')}\right)^{-1},$$ i.e.
\be\label{co}c^{(x_0'',u'')}_{(x_0',u')}+c^{(x_0',u')}_{(x_0,u)}=c^{(x_0'',u'')}_{(x_0,u)}
+\sum_k\left((w_{u'}')_k-\frac{t_0'}{2}v_k\right)v_k\,.\ee If we take another family of constants
$$\wt{c}^{(x_0'',u'')}_{(x_0',u')}=c^{(x_0'',u'')}_{(x_0',u')}+d^{(x_0'',u'')}_{(x_0',u')}\,,$$
then (\ref{co}) implies
\be\label{co1}d^{(x_0'',u'')}_{(x_0',u')}+d^{(x_0',u')}_{(x_0,u)}=d^{(x_0'',u'')}_{(x_0,u)}.\ee
But, as easily seen, the only functions $d$ on an affine
finite-dimensional space that satisfy (\ref{co1}) are of the form
$$d^{(x_0',u')}_{(x_0,u)}=A(x_0',u')-A(x_0,u)$$ for certain function $A$, i.e. we get only
the obvious freedom in constructing the line bundle. Thus we get
the following.
\begin{theorem}\label{t1} Let us fix a class of inertial observers $u\in E_1$. The transformations
(\ref{cc}) respect the Schr\"odinger operator (\ref{S1}) and
satisfy the cocycle condition (\ref{cc1}) if and only if the
functions $F^{(x_0'',u'')}_{(x_0',u')}$ are of the form
\be\label{ff} F^{(x_0'',u'')}_{(x_0',u')}(y,t)=
\frac{im}{\hbar}\left(\sum_k\left(y_k-\frac{t}{2}v_k'\right)v_k'+\sum_k\left((w_{u'}')_k-\frac{t_0'}{2}v_k\right)v_k\right)
+A(x_0'',u'')-A(x_0',u')\,,
\ee
for $w'_{u'}=y((x_0'-x_0'')-\zt(x_0'-x_0'')u')$ being the coordinates of $x_0'-x_0''\in V$ with respect to the
inertial observer $(x_0',u')$, for $t_0'=\zt(u'-u'')$, for $v'=(v'_k)$ being the coordinates of $u'-u''\in
E_0$, for $v=(v_k)$ being the coordinates of $u-u'\in E_0$, and $A$ being an arbitrary function $A:N\ti
E_1\ra\C$.
\end{theorem}
\begin{remark} The fact that certain transformations of the form (\ref{4a}) act on solutions of the
Schr\"odinger equation in different reference frames is known (see e.g. \cite[p.
100]{Pa} or \cite[section 4.3]{EMS}). Here, we have found a general form of such
transformations in order to recognize properly the arguments of the Schr\"odinger
operator. Moreover, such transformations have been proven to be unique up to the
obvious freedom.
\end{remark}
Removing constants from (\ref{ff}) we will stay in the same of projective class of the corresponding principal
$U(1)$ bundle. Thus we get the following.
\begin{theorem}\label{t2}
There is a unique projective class $\bP_m$ of principal $U(1)$-bundles $P_m$ over the Newtonian space-time
with trivializations $\Psi_{(x_0,u)}:P_m\ra\R^3\ti\R\ti U(1)$ indexed by inertial observers $(x_0,u)\in N\ti
E_1$ and covering the coordinate maps on the base
$$(y,t)(x)=\zf_{(x_0,u)}(x)=(y(x-x_0-\zt(x-x_0)u),\zt(x-x_0))$$
such that the transition maps
$$T^{(x'_0,u')}_{(x_0,u)}=\Psi_{(x'_0,u')}\circ\left(\Psi_{(x_0,u)}\right)^{-1}:\R^3\ti\R\ti U(1)\ra\R^3\ti\R\ti U(1)$$
leave the Schr\"odinger operator $\cS^0_m$ invariant. This projective class is represented by the projective
cocycle
\be\label{pc}\cT^{(x'_0,u')}_{(x_0,u)}(y,t,z)=\left(y+v(t+t_0)+w_u,t+t_0,e^{\frac{im}{\hbar}\left(\la
y,v\ran+\frac{t}{2}\Vert v\Vert^2\right)}\cdot z\right)\,,\ee
where $v\in\R^3$ are coordinates of $u-u'\in E_0$ and
$(w_u,t_0)=\left(y(x_0-x_0'-\zt(x_0-x_0')u),\zt(x_0-x_0')\right)$
are coordinates of $x_0-x'_0$ for any inertial observer $(x_0,u)$
in the class of $u$.
\end{theorem}
\noindent Any representative of the class $\bP_m$ we call a {\it
Schr\"odinger principal bundle} and the corresponding complex line
bundle $L_m$ -- the {\it Schr\"odinger line bundle}.

According to Theorem \ref{t1}, the differential operator
$\bcS^{(x_0',u')}_m$ on $L_m$, that corresponds to $\cS^0_m$ on
the trivial 1-dimensional vector bundle $\R^3\ti\R\ti\C$ {\em via}
the trivialization $\Psi_{(x_0',u')}$, does not depend on the
trivialization, so it gives rise to a well-defined differential
operator $\bcS^0_m$ on $L_m$. Choosing a potential $U\in
C^\infty_\C(N)$ we can write the full Schr\"odinger operator as
$\bcS^U_m\zc=\bcS^0_m\zc+U\zc$ acting on sections of $L_m$. We can
summarize these observations as follows.
\begin{theorem}
For any function (potential) $U$ on the Newtonian space-time $N$
there is a well-defined (trivialization-independent) differential
operator $\bcS^U_m$ ({\em the Schr\"odinger operator}), acting on
sections of the Schr\"odinger line bundle $L_m$. This operator
corresponds, {\em via} the trivialization $\Psi_{(x_0,u)}$, to the
differential operator \be\label{Ss}
\cS_U^m\zc=\frac{\hbar^2}{2m}\sum_k\frac{\pa^2\zc}{\pa
y_i^2}+i\hbar\frac{\pa\zc}{\pa t}-(U\circ\zf_{(x_0,u)}^{-1})\zc
\ee acting on complex functions $\zc(y,t)$ on $\R^3\ti\R$.
\end{theorem}
\noindent A Schr\"odinger principal bundle $P_m$ can be, for
example, constructed according to the general scheme
(\ref{construction}). Let us fix $u\in E_1$ and put in (\ref{ff})
$A=0$. Then, the transition maps corresponding to the phase change
$F$ can be written in the form
\bea\label{trm}
&T^{(x''_0,u'')}_{(x'_0,u')}(y,t,z)=\left(\zvy^{(x''_0,u'')}_{(x'_0,u')}\left(y,t\right),
e^{F^{(x''_0,u'')}_{(x'_0,u')}\left(\zvy^{(x''_0,u'')}_{(x'_0,u')}\left(y,t\right)\right)}\cdot z\right)\\
&=\left(y+w'_{u'}+ (t+t_0')v', t+t_0'\,,\exp{\left(\frac{im}{\hbar}\left(\left\la
y+w'_{u'}+ \frac{1}{2}(t+t_0')v',v'\right\ran+\left\la
w'_{u'}+\frac{t_0'}{2}v,v\right\ran \right)\right)}\cdot z\right)\,,\nn
\eea
where $w'_{u'}=y((x_0'-x_0'')-\zt(x_0'-x_0'')u')$, $t_0'=\zt(u'-u'')$, $v'=y(u'-u'')$, and $v=y(u-u')$. The
set $P_m^u$ of equivalence classes of the relation:
\be\label{eq}
(x_0',u', y',t', z')\sim (x_0'',u'', y'',t'',
z'')\quad\Longleftrightarrow\quad
T^{(x''_0,u'')}_{(x'_0,u')}(y',t',z')=(y'',t'',z'')\,.
\ee
defined on the product $N\times E_1\times \R^3\times \R\times
U(1)$ is a principal $U(1)$-bundle over $N$ with the projection
$$[x'_0,u',y',t',z']\longmapsto x'_0+y^{-1}(y')+t'u'=\zf^{-1}_{(x'_0,u')}(y',t')\in N\,.$$
For each inertial observer $(x'_0,u')$ in each equivalence class
of the relation $\sim$ there is one representative with
$(x'_0,u')$ in the first two places. It means that we have a
mapping
$$\Psi_{(x'_0,u')}: P_m^u\ni [x'_0,u',y',t',z']\longmapsto (y',t',z')\in (\R^3\ti\R\times U(1))$$
which is the trivialization (over $\R^3\ti\R$) of $P_m^u$
corresponding to the inertial observer $(x'_0,u')$ and
$$\Psi_{(x''_0,u'')}\circ\Psi_{(x'_0,u')}^{-1}=T^{(x''_0,u'')}_{(x'_0,u')}\,,$$
so the pair $(P_m^u, \Psi)$ is a principal bundle with trivialization which is a
representative of the class $\bP_m$.
\begin{remark}Of course, as solving a concrete Schr\"odinger  equation always takes place in a given
coordinate system, introducing the concept of the Schr\"odinger bundle does not imply
new methods in finding the solutions. It just gives a geometrical structure capturing
the necessary gauging of the wave functions while passing from one inertial frame to
another. All the geometrical setting supports the idea that wave functions should be
understood as classes $[\zc]$ not feeling a change by a constant phase. On the
principal Schr\"odinger bundle such a class is represented by an invariant horizontal
foliation, so by a flat principal connection. It is interesting that in this setting,
one can associate with a class of inertial observers moving with velocity $v$ with
respect to a given one a plane wave
$$W_v(y,t)=\exp\left[\frac{im}{\hbar}\left( \sum_kv_ky_k-\frac{t}{2}\sum_kv_k^2\right)\right]\,.$$
We should multiply a wave function by this plane wave, so change its phase by the phase
of this plane wave, before writing the wave functions in coordinates associated with
the new observer. In this sense, for quantum systems, different inertial observers
carry not only relative velocities but also relative plane waves.
\end{remark}
\section{Relation to Newtonian mechanics}
By means of a group homomorphism
     \be\label{u2} \R  \rightarrow U(1)\colon  s \mapsto \exp\left(\frac{i s}{\hbar}\right), \ee
     the Schr\"odinger principal $U(1)$-bundle $P_m$ can be considered as the reduced
     principal $(\R,+)$-bundle $\Z_m$ and the "additive projective class" of $\Z_m$ does not
     depend on the choice of $P_m$. For direct calculation we can use the bundle $\Z_m^u$
     - the "logarithm" of $P_m^u$ with trivializations transforming according to
\bea\label{trmav}
&\bar{T}^{(x''_0,u'')}_{(x'_0,u')}(y,t,s)=\\&\left(y+w'_{u'}+ (t+t_0')v',
t+t_0'\,,s+m\left\la y+w'_{u'}+ \frac{1}{2}(t+t_0')v',v'\right\ran+m\left\la
w'_{u'}+\frac{t_0'}{2}v,v\right\ran \right)\,.\nn
\eea
It is an AV-bundle in terminology of  \cite{GGU2}. Analogously as in (\ref{eq}), an
element of \ $\Z_m^u$ is an equivalence class of $(x_0',u', y',t', s')\in N\ti
E_1\ti\R^3\ti\R\ti\R$
\be\label{eqav} (x_0',u', y',t', s')\sim (x_0'',u'', y'',t'', s'')\quad\Longleftrightarrow\quad
\bar{T}^{(x''_0,u'')}_{(x'_0,u')}(y',t',s')=(y'',t'',s'')\,,
\ee
and the projection $\zz:\Z_m^u\ra N$ on $N$ reads
$$[x'_0,u',y',t',s']\longmapsto x'_0+y^{-1}(y')+t'u'=\zf^{-1}_{(x'_0,u')}(y',t')\in N\,.$$
Since $N$ is fibred over affine time, $\bar{\zt}:N\ra\T$, the standard construction of
the Hamiltonian AV-bundle \cite{GGU2,GU,U} yields
$$\Ph_\zz:\Ph(\Z_m^u)\ra\ul{\Ph(\Z_m^u)}\,,
$$
where $\Ph(\Z_m^u)$ is the phase bundle of the AV-bundle $\Z_m^u$ and
$$\ul{\Ph(\Z_m^u)}=\Ph(\Z_m^u)/\la\xd t\ran$$
(see \cite{GGU2,GGU3,GU,U}). Using a trivialization we can identify the above fibration
with
$$\Ph_\zz:\sT^*(\R^3\ti\R)\ra(\sT^\ast \R^3\ti\R)/\la\xd t\ran\,.$$
The transition maps (\ref{trmav}) act on sections $\zs$, represented in the
trivializations by functions $\zs=\zs(y,t)$, as
$$\left(\bar{T}^{(x''_0,u'')}_{(x'_0,u')}\right)_\ast(\zs)(y,t)=
\zs(y-tv'-w'_{u'},t-t_0')+m\left(\la y-\frac{t}{2}v',v'\ran+\la
w'_{u'}-\frac{t_0'}{2}v,v\ran\right)\,,$$ so the adapted Darboux coordinates in
$\sT^*(\R^3\ti\R)$ transform according to
\be\label{hamav}\Ph\left(\bar{T}^{(x''_0,u'')}_{(x'_0,u')}\right)(y,t,p_y,p_t)=
\left(y+w'_{u'}+ (t+t_0')v', t+t_0',p_y+mv',p_t-\la p_y,v'\ran-\frac{m}{2}\Vert
v'\Vert^2\right)\,.
\ee
Since, by convention, the distinguished vertical vector field on the hamiltonian AV-bundle is $-\pa_{p_t}$,
the vertical coordinate -- value of Hamiltonian sections -- is $h=-p_t$ and in coodinates $(y,t,p_y,h)$ we get
$\Ph_\zz(y,t,p_y,h)=(y,t,p_y)$, and the transition maps in the form
$$\Ph\left(\bar{T}^{(x''_0,u'')}_{(x'_0,u')}\right)(y,t,p_y,h)=
\left(y+w'_{u'}+ (t+t_0')v', t+t_0',p_y+mv',h+\la p_y,v'\ran+\frac{m}{2}\Vert
v'\Vert^2\right)\,.
$$
Note that these transformations do not depend on the distinguished
$u\in E_1$ nor $x_0',x_0''$ any longer but only on the relative
velocity $v'=u'-u''$, so the Hamiltonian bundle
$\bH_m=\Ph(\Z_m^u)$ does not depend on $u$ and in fact on the
choice of $P_m$ in the projective class $\bP_m$. In this bundle,
during transitions, the momenta (as elements of $E_0^\ast$)
transform according to the rule $p\mapsto p+m\la v',\cdot\ran$,
and the values of possible Hamiltonian sections -- according to
the rule $h\mapsto h+\la p,v'\ran+\frac{m}{2}\Vert v'\Vert^2$,
which is precisely the transformation used in \cite{GU,GGU2} to
define the Hamiltonian AV-bundle for a Newtonian particle of mass
$m$. This means that the AV-bundle $\Z_m^u$ plays the role of the
affine Hamilton-Jacobi bundle: the Hamilton-Jacobi equation is an
equation of sections $\zs$ of $\Z_m^u$. This bundle, however, is
not uniquely determined. If $\bd\zs:N\ra\Ph(\Z_m^u)=\bH_m$ denotes
the affine de Rham differential, then the Hamilton-Jacobi equation
associated with the Hamiltonian section $h:\ul{\bH_m}\ra{\bH_m}$
takes the form
$$\bd\zs(N)\subset h(\ul{\bH_m})\,.$$
In coordinates, this Hamilton-Jacobi equation takes the standard form
$$h \left(y,t,\frac{\pa\zs}{\pa y}\right)+\frac{\pa\zs}{\pa t}(y,t)=0\,.$$

\section{Atiyah bundle and generalized differential calculi}
Let us fix a principal Schr\"odinger bundle $P_m$. If we use the parametrization
\be\label{u1}\R\ni r\mapsto \exp\left(-\frac{im }{\hbar}r\right)\in U(1)\ee of $U(1)$,
then the change of coordinates (\ref{trm}) in $P_m$ associated with the change of
inertial frames reads
\bea\label{trm1}
&T^{(x''_0,u'')}_{(x'_0,u')}(y,t,r)=\left(\zvy^{(x''_0,u'')}_{(x'_0,u')}\left(y,t\right),
r-\frac{\hbar}{im}F^{(x''_0,u'')}_{(x'_0,u')}\left(\zvy^{(x''_0,u'')}_{(x'_0,u')}\left(y,t\right)\right)\right)\\
&=\left(y+w'_{u'}+ (t+t_0')v', t+t_0'\,,r-\left\la y+w'_{u'}+
\frac{1}{2}(t+t_0')v',v'\right\ran-\left\la w'_{u'}+\frac{t_0'}{2}v,v\right\ran
\right)\,.\nn
\eea
Let us observe now that every smooth section $\zc:N\ra P_m$ gives rise to a smooth complex function $\wt{\zc}$
on $P_m$ defined by
\be\label{ext}\wt{\zc}\left(\exp\left(\frac{im }{\hbar}r\right)\cdot\zc(x)\right)=
\exp\left(\frac{im }{\hbar}r\right)\,,\ee for any $n\in N$. In
coordinates associated with a choice of an inertial frame,
\be\label{psiext}\wt{\zc}(y,t,r)=e^{\frac{imr}{\hbar}}\zc(y,t).\ee
We can use the same local formula to produce the function
$\wt{\zc}$ on $P_m$ also from a section $\zc$ of the Schr\"odinger
complex line bundle $L_m$ associated with $P_m$:
$$\wt{\zc}\left(\exp\left(\frac{im }{\hbar}r\right)\cdot\frac{\zc(x)}{\vert \zc(x)\vert}\right)=
\exp\left(\frac{im }{\hbar}r\right)\vert\zc(x)\vert\,,$$ if $\zc(x)\ne 0$, and
$\wt{\zc}=0$ on the fibre over $x$ otherwise. Note that the "absolute value"
$\vert\zc(x)\vert$ is well defined on $L_m$, since it is a complex line bundle
associated with an $U(1)$-principal bundle. Moreover, the principal bundle $P_m$ can be
considered to be the set of unitary elements of $L_m$.

The functions of the form $\wt{\zc}$ on $P_m$ are characterized as
$\frac{im}{\hbar}$-homogeneous functions with respect to the
fundamental vector field $\pa_r$ of the $U(1)$-action. Indeed, if
$\pa_r(f)=\frac{im}{\hbar}f$, then the function $\zc$ written in
our coordinates as $e^{-\frac{imr}{\hbar}}f$ represents a section
of the Schr\"odinger bundle $L_m$. To see this, note first of all
that $\zc=\zc(y,t)$ does not depend on $r$. Second, under the
change of coordinates (\ref{trm1})
$$f(y,t,r)=\zc(y,t)e^{\frac{imr}{\hbar}}$$ is pushed forward into
\bea\label{d}f\circ
\left(T^{(x''_0,u'')}_{(x'_0,u')}\right)^{-1}(y,t,r)&=&
f\left(\left(\zvy^{(x''_0,u'')}_{(x'_0,u')}\right)^{-1}\left(y,t\right),\,
r+\frac{\hbar}{im}F^{(x''_0,u'')}_{(x'_0,u')}(y,t)\right)\\
&=&\zc\circ\left(\zvy^{(x''_0,u'')}_{(x'_0,u')}\right)^{-1}\left(y,t\right)\cdot
e^{F^{(x''_0,u'')}_{(x'_0,u')}(y,t)}\cdot e^{\frac{imr}{\hbar}}\,.\nn
\eea
But (\ref{d}) is $\zc'(y,t)e^{\frac{imr}{\hbar}}$, where $\zc'$ is the push-forward of
$\zc$, so $\zc$ is pushed forward according to the rule
$$\zc\mapsto e^{F^{(x''_0,u'')}_{(x'_0,u')}}\cdot\zc\circ
\left(\zvy^{(x''_0,u'')}_{(x'_0,u')}\right)^{-1}\,,
$$
i.e. exactly like sections of the Schr\"odinger bundle do. Thus we get the following.
\begin{theorem}\label{te} The local formula $\wt{\zc}(y,t,r)=\zc(y,t)e^{\frac{imr}{\hbar}}$
establishes a one-to-one correspondence between sections $\zc$ of
the Schr\"odinger line bundle $L_m$ and
$\frac{im}{\hbar}$-homogeneous (with respect to the fundamental
vector field $\pa_r$) functions on $P_m$.
\end{theorem}

\begin{remark}The above correspondence between sections of the Schr\"odinger principal $U(1)$-bundle
$P_m$ and functions on $P_m$ is similar to the analogous correspondence between
sections of an AV-bundle $\bA$ and functions on $\bA$ as exploited in
\cite{GGU1}-\cite{GGU3}. The latter can be viewed as a `classical' counterpart of this
correspondence for $U(1)$-principal bundles with trivializations (see the next
section). The function $\wt\zc$ on $P_m$ obtained from a section $\zc$ of the bundle
$L_m$ coincides with a function on $P_m$ obtained from $\zc$ by viewing at the
associated line bundle $L_m$ as the reduced trivial bundle $\C \times P_m$.
\end{remark}
Let us consider now the complex {\em Atiyah bundle} $\A_m$ over $N$ associated with the
principal $U(1)$-bundle $P_m$. Let us also recall that the Atiyah bundle can be
characterized as the vector bundle over the base of the principal $G$-bundle $P$ whose
sections are represented by $G$-invariant vector fields on $P$. In our case we choose
vector fields with complex coefficients which makes no real difference. As such vector
fields are projectable, we have a canonical surjective bundle map $\zr:\A_m\ra\sT N$
(the {\em anchor map}) with the kernel $KP_m$. Moreover, since invariant vector fields
are closed with respect to the Lie bracket, we have a canonical {\em Lie algebroid}
structure on $\A_m$ - the {\em Atiyah Lie algebroid} of $P_m$. For detailed description
of Lie and Atiyah algebroids we refer to the monograph \cite[Section 3.2]{Mac}. The
sections of $\A_m$ in our case are represented by complex vector fields on $P_m$,
commuting with the fundamental vector field $\pa_r$ of the $U(1)$-action and the Lie
algebroid bracket is represented by the commutator of vector fields. In coordinates
associated with a trivialization of $P_m$ they are of the form
\be\label{sec}X=\sum_kf_k(y,t)\pa_{y_k}+g(y,t)\pa_t+h(y,t)\pa_r\,.\ee
Every such invariant vector field -- section of $\A_m$ -- can be
canonically interpreted, in turn, as a first-order differential
operator $D_{X}$ on the Schr\"odinger complex line bundle. Indeed,
as such a vector field commutes with $\pa_r$, it acts on
$\frac{im}{\hbar}$-homogeneous functions $\wt{\zc}$, so sections
$\zc$ of $L_m$ by
$$\wt{(D_{X}(\zc))}=X(\wt{\zc}).$$
Since
$$X(\wt{\zc})=X(\zc\cdot e^{\frac{imr}{\hbar}})=\left(\sum_kf_k(y,t)
\frac{\pa\zc}{\pa {y_k}}(y,t)+g(y,t)\frac{\pa\zc}{\pa t}(y,t)
+\frac{im}{\hbar}h(y,t)\zc(y,t)\right)e^{\frac{imr}{\hbar}}\,,$$ the section
(\ref{sec}) of $\A_m$ represents in coordinates the first-order differential operator
\be\label{do}D_{X}=\sum_kf_k(y,t)\pa_{y_k}+g(y,t)\pa_t+\frac{im}{\hbar}h(y,t)\ee
acting on sections of the Schr\"odinger complex line bundle $L_m$. It is easy to see that the Lie algebroid
structure on $\A_m$ is represented by the standard commutator of differential operators. Note however that, as
there is no canonical trivialization of $L_m$, the space of sections does not carry a canonical structure of
an associative algebra, so derivations are not distinguished. We will call the sections of $\A_m$ -- {\it
Schr\"odinger vector fields}. In general, tensor fields built out of $\A_m$ we will call {\it Schr\"odinger
tensor fields}. They are represented by invariant tensor fields on the Schr\"odinger principal bundle $P_m$.
In particular, {\it Schr\"odinger $k$-forms} are sections of $\bigwedge^k\A_m^\ast$ and they are represented
by $U(1)$-invariant $k$-forms on $P_m$. However, if for a given trivialization $\Psi_{(x_0,u)}$ we interpret
the functional coefficients of a tensor field as {\it wave functions} -- sections of $L_m$ -- we get {\it wave
tensor fields}, i.e. sections of the corresponding tensor bundle of $\A_m$ tensored (over $N$) with $L_m$. In
particular, {\it wave functions} are sections of $L_m$, {\it wave forms} are sections of
$\left(\bigwedge^k\A_m^\ast\right)\ot_NL_m$, and {\it wave-vector fields} are sections of $\A_m\ot L_m$. Under
transition maps, the wave-tensor fields transform with a change in phases exactly like wave-functions.
\smallskip\noindent
We can extend the observation of Theorem \ref{te} to wave-tensor fields.
\begin{theorem}The formula $\wt{\zw}(y,t,r)=\zw(y,t)e^{\frac{imr}{\hbar}}$,
expressed in coordinates associated with a distinguished trivialization
$\Psi_{(x_0,u)}$, establishes a one-to-one correspondence between wave-tensor fields
$\zw$ and $\frac{im}{\hbar}$-homogeneous (with respect to the fundamental vector field
$\pa_r$) tensor fields $\wt{\zw}$ on the Schr\"odinger principal bundle $P_m$. This
correspondence depends on the trivialization.
\end{theorem}
\noindent On the wave-forms we have an analog $\wt{\xd}$ of the standard de Rham differential $\xd$, defined
by
$$\widetilde{(\wt{\xd}\zw)}=\xd\wt{\zw}.$$
Of course, by definition, $\wt{\xd}^2=0$. In coordinates associated with a choice of an inertial frame, this
differential reads \be\label{dR} \wt{\xd}\zw=\xd\zw+\frac{im}{\hbar}\xd r\we\zw.\ee We will call it {\em
wave-de Rham differential}. We hope this explanations makes clear that the contraction of a wave-vector field
with a $k$-covariant Schr\"odinger tensor is a $(k-1)$-covariant wave-tensor, as the contraction of a
$\frac{imr}{\hbar}$ -- homogeneous vector field with an $U(1)$-invariant $k$-covariant tensor is a
$\frac{imr}{\hbar}$-homogeneous covariant $(k-1)$-tensor.

\begin{remark}The local form of the wave-de Rham differential is a particular case of a deformation
of the de Rham differential considered already by E.~Witten \cite{Wi},
$\wt{\xd}\zw=e^{-\frac{imr}{\hbar}}\cdot\xd\left(e^{\frac{imr}{\hbar}}\zw\right)$ and
generalized to {\em Jacobi algebroids} (generalized Lie algebroids) in \cite{IM,GM1,
GM2}.
\end{remark}
\section{Schr\"odinger metrics}
Consider now a pseudo-Riemannian metric $\zm_m^{(x_0,u)}\in\Sec(\A_m^\ast\ot\A_m^\ast)$ on the Schr\"odinger
principal bundle $P_m$ such that $\zm_m^{(x_0,u)}$ corresponds {\it via} the trivialization
$\Psi_{(x_0,u)}:P_m\ra\R^3\ti\R\ti U(1)$ (associated with an inertial frame $(x_0,u)\in N\ti E_1$) to a
pseudo-Riemannian $U(1)$-invariant metric $\zm$ on $\R^3\ti\R\ti U(1)$ which extends the standard spatial
Euclidean metric on $\R^3\ti\R$, i.e. to a metric $\zm$ of the form
\bea\nn\zm(y,t,r)&=&\sum_k\xd y_k\ot\xd y_k+
\sum_kB_k(y,t)\xd y_k\vee\xd r\\&&+ C(y,t)\xd r\ot\xd r+D(y,t)\xd t\vee\xd
r\,.\label{metricc}
\eea

If we assume additionally that $\zm$ is invariant with respect to the change of
coordinates (\ref{trm1}), then $\zm_m=\zm^{(x_0,u)}_m$ is a pseudo-Riemannian metric on
$P_m$ which does not depend on the choice of trivialization. Such metric $\zm_m$ we
will call {\it Schr\"odinger metric}. Since $\zm$ is $U(1)$-invariant, looking for
Schr\"odinger metrics, we can forget about shifts in the coordinate $r$ and look for
$\zm$ which is invariant with respect to all maps
$$(y,t,r)\mapsto\left(y+(t+t_0)v+w,\,t+t_0,\,
r-\sum_kv_k(y_k+\frac{t}{2}v_k)\right).
$$
Straightforward calculations show that $B_k$ and $C$ must be 0, and $D=1$. Thus we get
the following.
\begin{theorem}\label{metric} There is a unique Schr\"odinger metric $\zm_m$ on $P_m$.
In coordinates associated with any bundle trivializations $\Psi_{(x_0,u)}$, $\zm_m$ it is given by
\be\label{SM} \zm_m=\sum_k\xd y_k\ot\xd y_k+
\left(\xd t\ot\xd r+\xd r\ot\xd t\right)\,.\ee
\end{theorem}
\noindent It is easy to see that the contravariant form of the Schr\"odinger metric $\zm_m$ in coordinates
reads
\be\label{SMC}
\zn_n=\sum_k\pa_{y_k}\ot\pa_{y_k}+ \left(\pa_t\ot\pa_r+\pa_r\ot\pa_t\right)\,.
\ee
\noindent A $\zn$-orhogonal basis of 1-forms is for example $\xd{y_k},\zb_+,\zb_-$, where $\xd{y_k}$ and
$\zb_+=\frac{\xd r+\xd t}{\sqrt{2}}$ have length 1 and $\zb_-=\frac{\xd r-\xd t}{\sqrt{2}}$ has squared length
$-1$. Therefore, the {\em Schr\"odinger volume} $\zW_m$ associated with the Schr\"odinger metric $\zm_m$ (and
defined up to a sign) is represented by
\be\label{volume}\zW_m=\xd
y\we\zb_+\we\zb_-= \xd y\we \xd t\we \xd r\,,\ee where $\xd y=\xd y_1\we\xd y_2\we\xd y_3$.

\begin{remark}The metric $\zm$ can be transported to a metric on the total space of a Hamilton-Jacobi
bundle $\Z_m^u$. The total space of $\Z_m^u$ is an affine space and, for $m=1$, the
metric satisfies the properties of a Galilei metrics postulated in \cite{WMT3}. Thus
$\Z_1^u$ is an example of a Galilei space. A wave function on Galilei space (without
potential) satisfies the Laplace equation for the Galilei metric and is
$\frac{im}{\hbar}$-homogeneous. This shows full compatibility of our four-dimensional
approach with the wave mechanics of the Galilei space.
\end{remark}
\section{Schr\"odinger-Laplace operators for the Schr\"odinger metrics}
With the use of the Schr\"odinger differential $\wt{\xd}$ and the Schr\"odinger metric $\zm_m$ one can define
the {\em wave-gradient} $\nabla_\zc$ of a wave-function $\zc$ -- a section of the Schr\"odinger complex line

bundle $L_m$ -- in the standard way:
\be\label{grad}i_{\nabla_\zc}\zm_m={\wt{\xd}\zc}.\ee
The wave-gradient is clearly a wave-vector field. In coordinates,
$$\wt{\xd}\zc=\sum_k\frac{\pa\zc}{\pa y_k}\xd y_k+\frac{\pa\zc}{\pa t}\xd t+
\frac{im}{\hbar}\zc\xd r$$ and
$$\nabla_\zc=\sum_k\frac{\pa\zc}{\pa y_k}\pa_{y_k}+\frac{im}{\hbar}\zc\pa_t+
+\frac{\pa\zc}{\pa t}\pa_r\,,
$$
where the functional coefficiants should be understood as wave-functions.

For every wave-vector field $Y$, in turn, its {\em wave-divergence} $\text{div}(Y)$ -- associated with the
Schr\"odinger metric $\zm_m$ -- is defined via the Schr\"odinger volume $\zW_m$, like classically, as
\be\label{div} {\text{div}(Y)\zW_m}=\wt{\xd}(i_Y\zW_m).
\ee
Here, $i_Y\zW_m$, thus $\wt{\xd}(i_Y\zW_m)$ is a wave-form, as well as the obviously
defined product of the wave-function $\text{div}(Y)$ and the Schr\"odinger volume form
$\zW_m$. In coordinates,
$$\text{div}(\sum_kf_k\pa_{y_k}+g\pa_t+h\pa_r)=
\sum_k\frac{\pa f_k}{\pa y_k}+\frac{\pa g}{\pa t}+\frac{im}{\hbar}h.
$$
And finally, we can define the {\em Schr\"odinger-Laplace operator} $\zD_m$, associated with the Schr\"odinger
metric $\zm_m$, by the formula completely analogous to the formula defining standard Laplace-Beltrami
operators:
\be\label{laplace} \zD_m\zc=\text{div}(\nabla_\zc).\ee The Schr\"odinger-Laplace operator is therefore a
second-order differential operators acting on the Schr\"odinger complex line bundle $L_m$, i.e. mapping wave
functions into wave functions. The above definition is completely intrinsic and natural. In coordinates
associated with a choice of an inertial frame,
\be\label{laplace-coordinates}
\zD_m\zc=\sum_k\frac{\pa^2\zc}{\pa{y_k}^2}+\frac{2im}{\hbar}\frac{\pa\zc}{\pa t}.\ee
But this is exactly the free Schr\"odinger operator $\bcS^0_m$ on $L_M$ up to a
constant factor:
$$\bcS^0_m\zc=\frac{\hbar^2}{2m}\zD_m\zc=\frac{\hbar^2}{2m}\sum_k\frac{\pa^2\zc}{\pa{y_k}^2}
+i{\hbar}\frac{\pa\zc}{\pa t}\,.$$

\medskip\noindent
\begin{example} Consider for simplicity $1+1$ dimensional space-time and inertial frames
differing only by the relative velocity $v\in\R$. For fixed mass $m>0$, with the
relative velocity $v$ we associate the plane wave $W_v(y,t)$ on $\R\ti\R$ with
coordinates $(y,t)$ by
$$W_v(y,t)=\exp\left[\frac{im}{\hbar}\left( yv-\frac{t}{2}v^2\right)\right]\,.$$
The Schr\"odinger line bundle $L_m$ in this setting can be interpreted as quotient
$\wt{L}/\sim_\zp$ of the trivial complex line bundle $\wt{L}=E_1\ti\R\ti\R\ti\C$, where
$E_1$ is the affine $\R$ (no 0 chosen), modulo the action of the additive group $\R$
acting on $\wt{L}$ by $\R\ni v\mapsto \zp_v$,
\be\label{raction}\zp_v(u,y,t,z)=(u+v,y+vt,t,[W_v(y+vt,t)]^{-1}\cdot z)\,.\ee
This line bundle is associated with the Schr\"odinger $U(1)$-principal bundle $P_m$
obtained as the quotient of the trivial $U(1)$-principal bundle
$\wt{P}=E_1\ti\R\ti\R\ti U(1)$ modulo the $\R$-action completely analogous to
(\ref{raction}). The sections $\zc$ of $L_m$ (resp., $P_m$) are therefore interpreted
as sections of $\wt{L}$ (resp., $\wt{P}$), $z=\zc(u,y,t)$, which are invariant with
respect to this $\R$-action. Hence, for fixed $u\in E_1$, they are viewed as
complex-valued (resp., $U(1)$-valued) functions on $\R\ti\R$. With a section of $L_m$
represented by $z=\zc(u,y,t)$ we associate the function $\wt{\zc}$ on $P_m$ represented
by function $\wt{\zc}(u,y,t,z)=z\cdot{\zc}(u,y,t)$ on $\wt{P}$ which is simultaneously
$\R$-invariant and $U(1)$-invariant. Conversely, every bi-invariant complex-valued
function on $\wt{P}$ represents a section of $L_m$ in the above way. The differential
operator
$$\wt{D}_m=\pa_y^2+\frac{2im}{\hbar}\pa_t$$ is clearly $U(1)$-invariant. It is
also, $\R$-invariant, $\wt{D}_m(f\circ\zp_v)=\wt{D}_m(f)\circ\zp_v$ (what is less
trivial but straightforward), so it induces a frame-independent differential operator
$D_m$ on sections of $L_m$. When fixing $u\in E_1$, we get the standard free
Schr\"odinger operator
$$\bcS^0_m\zc=\frac{\hbar^2}{2m}D_m\zc=\frac{\hbar^2}{2m}\frac{\pa^2\zc}{\pa{y}^2}
+i{\hbar}\frac{\pa\zc}{\pa t}\,.$$ But the differential operator $\wt{D}_m$ acts on
$U(1)$-invariant functions $\wt{\zc}(u,y,t,z)=z\cdot{\zc}(u,y,t)$ on $\wt{P}$ as the
operator
$$\wt{\zD}_m=\pa_y^2+\frac{2imz}{\hbar}\pa_t\pa_z$$
which is the Laplace-Beltrami operator of the pseudo-Riemannian metric $\zm_m$
represented by the bi-invariant symmetric form
$$\zm=\xd y\ot\xd y+\frac{i\hbar\ol{z}}{m}(\xd t\ot\xd
z+\xd z\ot\xd t).$$ The operator $\wt{\zD}_m$ is the extended Schr\"odinger operator in
the sense of Lizzi-Marmo-Sparano-Vinogradov \cite{LMSV}.
\end{example}

\section{Concluding remarks}
We have found a proper geometrical setting for frame-independent understanding of the classical Schr\"odinger
operators on the Newtonian space-time and we have found a description of the free Schr\"odinger operator as a
(generalized) Laplace-Beltrami operator.

In this picture, the Schr\"odinger operators act not on functions on the space-time but on sections of certain
one-dimensional complex vector bundle -- {\em Schr\"odinger line bundle}. This line bundle has trivializations
indexed by inertial observers and is closely related to an $U(1)$-principal bundle with an analogous list of
trivializations -- {\em Schr\"odinger principal bundle}. If an inertial frame is fixed, the Schr\"odinger
bundle can be identified with the trivial bundle over space-time, but as there is no canonical trivialization
(inertial frame) these sections, interpreted as wave-functions, cannot be viewed as actual functions on the
space-time. A change of an observer results not only in a change of coordinates  but also in the change of the
phase of the wave function.

The projective class of all possible Schr\"odinger bundles is
uniquely determined and its "logarithm" is an $\R$-principal
bundle whose sections are subject of Hamilton-Jacobi equations,
that makes a bridge between the classical and quantum theory.

On the {\em Schr\"odinger principal bundle} a natural (generalized) differential calculus is developed based
on a de Rham-like differential -- similar to the one considered by E.~Witten \cite{Wi} and similar to the
differential of so called {\em Jacobi algebroids} \cite{IM, GM1, GM2}. In this calculus, the (generalized)
Laplace-Beltrami operator associated with a naturally distinguished invariant pseudo-Riemannian metric on the
Schr\"odinger principal bundle turns out to coincide, up to a factor, with the classical free Schr\"odinger
operator.

The presented framework is conceptually four-dimensional (the base
is identified with the traditional Newtonian space-time but the
values of wave functions are not true numbers), does not involve
any {\em ad hoc} or axiomatically introduced geometrical
structures and it is based only on the traditional understanding
of the Schr\"odinger operator in a given reference frame. This
makes it mathematically simple, demonstrative, and respecting the
postulate of Occam's Razor.


\bigskip
\noindent Katarzyna Grabowska\\
Division of Mathematical Methods in Physics,
                University of Warsaw \\
                Ho\.za 69, 00-681 Warszawa, Poland \\
                 {\tt konieczn@fuw.edu.pl} \\\\
\noindent Janusz Grabowski\\Institute of Mathematics, Polish Academy of Sciences\\
\'Sniadeckich 8, P.O. Box 21, 00-956 Warszawa,
Poland\\{\tt jagrab@impan.gov.pl}\\\\
\noindent Pawe\l\ Urba\'nski\\
Division of Mathematical Methods in Physics,
                University of Warsaw \\
                Ho\.za 69, 00-681 Warszawa, Poland \\
                 {\tt urba\'nski@fuw.edu.pl}

\end{document}